\documentclass[letterpaper,aps,prd,twocolumn,tightenlines,preprintnumbers,nofootinbib,showkeys,superscriptaddress,longbibliography
]{revtex4-1}

\pdfoutput=1 
\usepackage[T1]{fontenc}

\usepackage{fullpage}
\usepackage{amsfonts}
\usepackage{amsmath}
\usepackage{slashed}
\usepackage{amssymb}
\usepackage{graphicx}
\usepackage{epic}
\usepackage{eepic}
\usepackage{epsfig}
\usepackage{latexsym}
\usepackage[dvipsnames]{xcolor}
\usepackage{float}
\usepackage{multirow}
\usepackage[breaklinks]{hyperref}
\usepackage{enumitem}
\hypersetup{colorlinks=true,citecolor=blue,linkcolor=blue,urlcolor=NavyBlue}
\usepackage[caption=false]{subfig}
\usepackage{natbib}
\usepackage{cancel}
\usepackage{relsize}
\usepackage[left=2cm,right=2cm,top=2cm,bottom=2cm]{geometry}
\usepackage{makecell}
\usepackage{soul}

\usepackage{bm}
\usepackage{subfig}

\usepackage{shorthand}
\usepackage{mciteplus}

\def\beq{\begin{equation}}
\def\eeq{\end{equation}}
\def\bea{\begin{eqnarray}}
\def\eea{\end{eqnarray}}

\def\h1{\ensuremath{h_1}}
\def\h2{\ensuremath{h_2}}

\newcommand{\hobs}{H_{\rm obs}}

\newcommand{\hpm}{H^\pm}

\begin{document}
\linespread{1.02}

\title{Electroweak Multi-Higgs Production: A Smoking Gun for the Type-I Two-Higgs-Doublet Model}


\author{Tanmoy Mondal}
\email{tanmoy@het.phys.sci.osaka-u.ac.jp}
\affiliation{Department of Physics, Osaka University, Toyonaka, Osaka 560-0043, Japan}
\affiliation{Birla Institute of Technology and Science, Pilani, 333031, Rajasthan, India}
\author{Stefano Moretti}
\email{s.moretti@soton.ac.uk; stefano.moretti@physics.uu.se}
\affiliation{School of Physics \& Astronomy, University of Southampton, Southampton SO17 1BJ, United Kingdom}
\affiliation{Department of Physics \& Astronomy, Uppsala University, Box 516, SE-751 20 Uppsala, Sweden}

\author{Shoaib Munir}
\email{munir1@stolaf.edu}
\affiliation{East African Institute for Fundamental Research (ICTP-EAIFR), University of Rwanda, Kigali, Rwanda}
\affiliation{Department of Physics, Faculty of Natural Sciences and Mathematics, St. Olaf College, Northfield, MN 55057, United States}

\author{Prasenjit Sanyal}
\email{prasenjit.sanyal01@gmail.com}
\affiliation{Department of Physics, Konkuk University, Seoul 05029, Republic of Korea}

\date{\today}

\begin{abstract}

Extending the Higgs sector of the Standard Model (SM) by just one additional Higgs doublet field leads to the two-Higgs-doublet model (2HDM). In the Type-I $Z_2$-symmetric limit of the 2HDM, all the five new physical Higgs states can be fairly light, $\mathcal{O}(100)$\,GeV or less, without being in conflict with  current data from the direct Higgs boson searches and the $B$-physics measurements. In this article, we establish that the new neutral as well as the charged Higgs bosons in this model can all be simultaneously observable in the multi-$b$ final state. The statistical significance of the signature for each of these Higgs states, resulting from the electro-weak (EW) production of their pairs, can exceed 5$\sigma$ at the 13\,TeV High-Luminosity Large Hadron Collider (HL-LHC). Since the parameter space configurations where this is achievable are precluded in the other, more extensively pursued, 2HDM Types, an experimental validation of our findings would be a clear indication that the true underlying Higgs sector in nature is the Type-I 2HDM.
\end{abstract}


\maketitle

\section{Introduction} \label{sec:intro}

The existence of additional Higgs bosons, besides the one discovered by the LHC \cite{Aad:2012tfa,Chatrchyan:2012xdj} (hereafter, denoted by $\hobs$), is predicted by most (if not all) frameworks of new physics. Observation of a second Higgs boson will thus provide firm evidence that the underlying manifestation of the EW Symmetry Breaking (EWSB) mechanism is a non-minimal one. 

From a theoretical point of view, given the fact that the $\hobs$ belongs to a complex doublet field 
in the SM, any additional Higgs field can be naturally expected to have the same $SU(2)_L$ representation. Following this argument, even the minimal bottom-up approach of augmenting the SM with a second doublet Higgs field and assuming CP-invariance yields a total of five physical Higgs states after EWSB: two neutral scalars ($h$ and $H$, with $m_h<m_H$), one pseudoscalar ($A$), and a charged pair ($\hpm$). If both the doublets $\Phi_1$ and $\Phi_2$ in this 2HDM couple to all the fermions of the SM, they would cause flavor-changing neutral currents (FCNCs) that contradict the experimental results. To prevent these FCNCs, a $\mathbb{Z}_2$ symmetry can be imposed \cite{Glashow:1976nt,Paschos:1976ay}, under which $\Phi_1 \to \Phi_1,\,\Phi_2 \to -\Phi_2, u_R^i\to -u_R^i,\, d_R^i\to -d_R^i,\,e_R^i\to -e_R^i$, so that all the quarks and charged leptons (conventionally) couple only to the $\Phi_2$, resulting in the so-called Type-I 2HDM (see \cite{Gunion:1989we,Branco:2011iw} for detailed reviews).

By now, many studies \cite{Arhrib:2009hc,Hespel:2014sla,Enberg:2016ygw,Arhrib:2016wpw,Arhrib:2017uon,Arhrib:2017wmo,Xie:2018yiv,Arhrib:2020tqk,Arhrib:2021xmc,Wang:2021pxc,Atkinson:2021eox,Mondal:2021bxa,Kanemura:2021dez,Krab:2022lih,Cheung:2022ndq,Kim:2022nmm,Kim:2023lxc,Chung:2022kjp} have established that the additional Higgs states (when the $\hobs$ is identified with either the $h$ or the $H$ state) of the 2HDM can be individually accessed at the LHC. Therefore, several searches for singly-produced neutral and charged Higgs bosons have been carried out by the ATLAS and CMS collaborations (see, e.g., \cite{Ivina:2022tfm,ATLAS:2022enb,ATLAS:2022vhr,ATLAS:2022eap,CMS:2022qww,CMS:2022suh,CMS:2022jqc,CMS:2022xxa}), but they remain elusive thus far. Even if a single state is eventually observed, the corresponding measurements that will ensue will, however, not enable one to ascertain which of the many possible extended realisations of the Higgs mechanism is at work. 

The majority of analyses, both phenomenological and experimental ones, involving an electrically neutral multi-Higgs final state, concentrate on QCD-induced production modes, namely, gluon-fusion and $b\bar b$-annihilation. While such gluon-initiated production is evidently highly dominant in the SM, it is not necessarily so in new physics models, owing to the non-standard couplings of their new Higgs bosons to the fermions and gauge bosons. In a previous analysis~\cite{Enberg:2018pye} it was shown that the inclusive cross sections for the $q\bar q^{(\prime)}$-induced production, where $q$ represents predominantly a $u$ or $d$ quark, of neutral multi-Higgs final states can be larger than their QCD-induced production, over sizeable parameter space regions of the Type-I 2HDM with standard hierarchy ($\hobs \equiv h$). The charged final states can of course only be produced via EW processes.

In this article, through a complete detector-level Monte Carlo (MC) analysis, we concretely establish that EW production can provide simultaneously visible signals of all the three additional Higgs bosons of the Type-I 2HDM at the LHC  with 3000\,fb$^{-1}$ integrated luminosity. The model parameter space configurations where this is possible contain an $A$ lighter than the $\hobs$, with the $H$ and $\hpm$ not much heavier, and are therefore well-motivated, in that the entire Higgs spectrum lies at the EW scale. Our signature channel, constituting of multiple $b$-quarks, allows a full reconstruction of the $H$, $A$ and $H^\pm$ masses. It implies that the LHC can uniquely pin down (or definitively rule out) the underlying EWSB mechanism as this (albeit narrow) parameter space region of the Type-I 2HDM (or at least as a low-energy manifestation of a grander framework with a Higgs sector mimicking this model). What makes our results all the more special is the fact that such a particular Higgs boson mass spectrum is forbidden in the Type-II 2HDM~\cite{Mahmoudi:2017mtv} (the realisation aligning with minimal Supersymmetry). 

The article is organised as follows. In Sec. \ref{sec:model} we very briefly review the Type-I 2HDM and its parameter space configurations relevant for multi-Higgs production, and identify a benchmark point (BP) satisfying the most important theoretical and experimental constraints. In Sec. \ref{sec:Signal} we detail our MC analysis, and in Sec. \ref{sec:signi} we establish the potential of the LHC to extract all the Higgs boson masses in the model. We present our conclusions in Sec. \ref{sec:summa}.


\section{The Type-I 2HDM}
\label{sec:model}

\subsection{Higgs potential and parameters}
\label{subsec:THDM}

The most general potential of a CP-conserving 2HDM can be written as
\begin{equation}
\begin{split}
\mathcal{V} &= m_{11}^2\Phi_1^\dagger\Phi_1+ m_{22}^2\Phi_2^\dagger\Phi_2
-[m_{12}^2\Phi_1^\dagger\Phi_2+ \, \text{h.c.} ] \\
& +\frac{\lambda_1}{2}(\Phi_1^\dagger\Phi_1)^2
+\frac{\lambda_2}{2}(\Phi_2^\dagger\Phi_2)^2
+\lambda_3(\Phi_1^\dagger\Phi_1)(\Phi_2^\dagger\Phi_2)\\
&+\lambda_4(\Phi_1^\dagger\Phi_2)(\Phi_2^\dagger\Phi_1) 
+\left[\frac{\lambda_5}{2}(\Phi_1^\dagger\Phi_2)^2
+\,\text{h.c.}\right]\,.
\label{eq:2hdmpot}
\end{split}
\end{equation}
It is convenient to write the doublets $\Phi_1$ and $\Phi_2$, after EWSB, in terms of their respective vacuum expectation values (VEVs) $v_1$ and $v_2$, the Goldstone bosons $G$ and $G^\pm$ and the physical Higgs states as
\begin{eqnarray}
\begin{split}
\Phi_1=\frac{1}{\sqrt{2}}\left(\begin{array}{c}
\displaystyle \sqrt{2}\left(G^+ c_\beta -H^+ s_\beta\right)  \\
\displaystyle v_1-h s_\alpha+H c_\alpha+i\left( G c_\beta-A s_\beta \right)
\end{array}
\right)\,,\\
\Phi_2=\frac{1}{\sqrt{2}}\left(\begin{array}{c}
\displaystyle \sqrt{2}\left(G^+ s_\beta +H^+c_\beta\right)  \\
\displaystyle v_2+h c_\alpha+Hs_\alpha+i\left( G s_\beta+A c_\beta \right)
\end{array}
\right)\,,
\end{split}
\end{eqnarray}
where $\beta$ $\equiv \tan^{-1}(v_2/v_1)$ and $\alpha$ are the angles rotating the CP-odd and the CP-even interaction states, respectively, into physical Higgs states, with $s_x$ ($c_x$) implying $\sin(x)$ ($\cos(x)$). Using the tadpole conditions of the $\mathcal{V}$, $m_{11}^2$ and $m_{22}^2$ can be replaced by $v_1$ and $v_2$ (and subsequently by $t_\beta$ -- short for $\tan\beta$ -- and $v\equiv\sqrt{v_1^2+v_2^2}=246$\,GeV) as the free parameters of the model. Furthermore, the physical Higgs boson masses and the parameter $s_{\beta-\alpha}$ can be traded in for $\lambda_{1-5}$. 

\subsection{Multi-$A$ production and benchmark scenarios}
\label{subsec:AA}

The benefit of using the physical Higgs boson masses as input parameters is that we can fix $m_h=125$\,GeV, so that our analysis corresponds to the `standard hierarchy' scenario with $h=\hobs$ and a heavier $H$. For this scenario, our previous study \cite{Enberg:2018pye} found that not only can the cross section for the EW production of the $HA$ pair be up to two orders of magnitude larger than the $gg/bb$-induced one, but it also remains quite substantial for the subsequent states $AAA$ and $AAZ$. Evidently, this cross section is more pronounced in parameter space regions where the $H$ is produced on-shell, with a mass just above the $AA$ or $AZ$ decay threshold and a maximal corresponding branching ratio (BR). The requirement of the couplings of the $h$ to be SM-like, as is the case for the $\hobs$, pushes the model into the so-called alignment limit, where $s_{\beta-\alpha}\to 1$ \cite{Bernon:2015wef}. In this limit, the $Hhh$ coupling is suppressed, unlike the $HAA$ coupling. The $HAZ$ coupling, which is proportional to $s_{\beta-\alpha}$, and hence the BR($H\to AZ$), is also naturally enhanced, while the $H\to VV$ decays, even when available, are suppressed. 

In light of the above observations, our analysis pertains to a small $m_A$ of 70\,GeV. For such a light $A$, $b\bar{b}$ is by far the dominant decay mode and the multi-Higgs states that we are interested in here are thus the ones yielding at least 4 $b$--quarks via intermediate $A$s. Such states result from the EW production of either a neutral pair of Higgs bosons, both on-shell, as
\begin{eqnarray}
AAA &:& q \bar{q} \to H (\to A A) A \to 4b+X\,, \nonumber \\
AAZ &:& q \bar{q} \to H (\to A Z) A \to 4b+X\,, \nonumber \\
AAWW &:& q \bar{q}\to H^+ (\to A W) H^-(\to A W)\to 4b+X\,, \nonumber
\nonumber   
\end{eqnarray}
or a charged pair, as
\begin{eqnarray}
AAW &:&  q \bar{q}^\prime \to H^\pm (\to A W) A \to 4b + X\,, \nonumber \\
AAAW &:& q \bar{q}^\prime \to H^\pm (H^\pm \to A W) H (\to A A) \to 4b+X\,, \nonumber \\  
AAZW &:& q \bar{q}^\prime \to H^\pm (H^\pm \to A W) H (\to A Z) \to 4b+X\,.\nonumber 
\end{eqnarray} 
Here the $W$ and $Z$ decay inclusively (i.e., both hadronically and leptonically) and $X$ can thus be any additional quarks (including $b$--quarks) and/or leptons. 

In order to find model configurations with substantial EW production cross sections for a representative value of $m_A=70$\,GeV, we numerically scanned the remaining parameters in the wide ranges
\begin{center}
$m_H$: $[2m_A-250]$\,GeV\,,~~$m_{\hpm}$: [100 -- 300]\,GeV\,,\\
$s_{\beta - \alpha}$: $0.9$ -- 1.0\,,~~$m_{12}^2$: 0 -- $m_A^2\sin\beta\cos\beta$\,,~~$t_\beta$: 1 -- 60\,, \\
\end{center} 
using the {\tt 2HDMC-1.8.0} code \cite{Eriksson:2009ws}. 
One of the most important constraints on the 2HDMs comes from the measurements of the oblique parameters $S$, $T$ and $U$, which in general forces $m_{\hpm}$ to lie close to $m_H$ and/or $m_A$. The {\tt 2HDMC} code internally calculates the theoretical predictions of these observables. In our scans, we required them to lie within the 95\% confidence level (CL) ellipsoid based on the 2022 PDG values \cite{ParticleDataGroup:2022pth}, $S = -0.01 \pm 0.07$ and $T = 0.04 \pm 0.06$, with correlations $\rho_{ST}=0.92$ for $U=0$. {\tt 2HDMC} also checks each scanned point against theoretical constraints such as vacuum stability, tree-level unitarity, and perturbativity ($|\lambda_i|<4\pi$). We moreover calculated the observable ${\rm BR}(B\to X_s \gamma)$ using the {\tt SuperIso-v4.1} \cite{Mahmoudi:2008tp} program, and ensured that its prediction lied outside the exclusion contour in the 
$\{m_{H^\pm}$,\,$t_\beta\}$ plane derived in~\cite{Mahmoudi:2017mtv,Sanyal:2019xcp} based on experimental results. 

Finally, we required all the Higgs states in each scanned point to satisfy the $95\%$ CL constraints included in {\tt HiggsBounds-v5.10.2} \cite{Bechtle:2013wla}. We additionally made sure that the SM couplings of the $h$ were consistent with the combined $2\sigma$ measurements for $\hobs$ from the ATLAS and CMS collaborations \cite{Langford:2021osp} using {\tt HiggsSignals-v2.6.2} \cite{Bechtle:2013xfa,Bechtle:2020uwn,Bahl:2022igd}. From the successfully scanned parameter space points, we extracted a BP, for which the BR($H\to AA$) is almost 1 (and hence BR$(H\to AZ)$ is strongly suppressed). Some specifics of this BP are given in Table~\ref{tab:params}. 

\section{Signal isolation}\label{sec:numerical}
\label{sec:Signal}

The background events for the multi-$b$ final states that we consider here
originate predominantly from the QCD multi-jet and $t \bar{t} + {\rm jets}$
processes. In our computation, we matched the multi-jet background up to four
jets and the $t\bar t$ up to two jets. Our matched cross section for the 
multi-jet background in the 5-flavor scheme at the LHC with $\sqrt{s}=13$\,
TeV is $8.98\times10^6$\,pb, with {\tt NNPDF23\_lo\_as\_0130}
\cite{NNPDF:2017mvq} parton distribution functions (PDFs) and a matching scale 
of 67.5\,GeV. The $t\bar t$ production cross section is 833.9\,pb, as calculated
with the {\tt Top++2.0} \cite{Czakon:2011xx} program, assuming a top quark mass
of 173.2\,GeV. For our simulation we generated $2\times 10^8$ multi-jet events and $10^7$ $t\bar t$ events. Other possible background processes include $t\bar{t}b\bar{b}$, $t\bar{t}+V$ (where $V=Z/W$), $V+{\rm jets}$, $ZZ$ and $hZ$, but we found them to be negligible after the selections. 
 
For our selected BP, we again used the {\tt NNPDF23\_lo\_as\_0130} PDF set to estimate the multi-$b$ signal events. We performed event-generation and parton shower with {\tt MadGraph5\_aMC@NLO}\cite{Alwall:2011uj,Alwall:2014hca} and 
{\tt Pythia-8.2} \cite{Sjostrand:2006za,Sjostrand:2014zea}, using the anti-$k_t$
algorithm \cite{Cacciari:2008gp} with $R=0.4$ for jet-reconstruction.
For $b$--tagging, we used the $p_T$-dependent efficiencies corresponding to the `DeepCSV Medium' working point based on the $\sqrt{s}=13$\,TeV data from the CMS collaboration \cite{CMS:2017wtu}. We used {\tt Delphes-3.4.2} \cite{deFavereau:2013fsa} for event generation, which was followed by analysis in the {\tt ROOT}~\cite{Brun:1997pa} framework.\footnote{We retained the default CMS jet energy scale in {\tt Delphes}.} The primary selection cuts we applied for signal
isolation include: $p_T > 20$\,GeV and $|\eta| < 2.5$ for all the jets in any
reconstructed object. Further selections that we made for each Higgs state are explained below. 

\subsection{Reconstruction of the $A$} 

\begin{enumerate}[leftmargin=*] 
\itemsep-0.05cm

\item Since all the signal processes contain at least two $A$s, the events should contain at least 4 $b$--jets, $a$, $b$, $c$ and $d$, which can be resolved into pairs 1 and 2. For this purpose we used a pairing algorithm for the leading $b$--jets to choose one combination out of the possible three: ($a,b\,;\;c,d)$, $(a,c\,;\; b,d)$, and $(a,d\,;\; b,c)$, which minimizes \cite{CMS:2022usq}
\begin{eqnarray}
\Delta R = |(\Delta R_1 - 0.8)| + |(\Delta R_2 - 0.8)| \,.
\end{eqnarray}  
Here, for a given combination, 
\begin{eqnarray}
\begin{split}
\Delta R_1 = \sqrt{(\eta_a - \eta_b)^2 + (\phi_a - \phi_b)^2}\,,\\
\Delta R_2 = \sqrt{(\eta_c - \eta_d)^2 + (\phi_c - \phi_d)^2}\,,
\end{split}
\end{eqnarray}
and offsetting each of these by 0.8 omits the $b$--jet pairings with too large an overlap in the $\{\eta,\,\phi\}$ space. This algorithm is motivated by the idea that the $b$--jets coming from a resonance (presumably the $A$) are closer together compared to the uncorrelated ones.

\item After the pairing, we imposed an asymmetry cut,
\begin{equation}\label{eq:alpha}
\bar{\alpha} =  \frac{|m_1 - m_2|}{m_1 + m_2} < 0.2\,,
\end{equation}
where $m_1$ and $m_2$ are the invariant masses of the two $b$--jet pairs. This cut ensures that these two pairs are from identical resonances, i.e., from $AA$. 

\end{enumerate}

\subsection{Reconstruction of the $H^\pm$} 
\begin{enumerate}[leftmargin=*] 
\itemsep-0.05cm

\item All events should contain at least 4 $b$--tagged jets and a pair of leading jets (thus corresponding to the dominant $q \bar{q}^\prime \to A_1 H^\pm \to A_1\,A_2\,W^\pm \to 4b+jj$ process).

\item The invariant mass of the leading $jj$ should lie within the $m_{W} \pm 25$\,GeV mass window.

\item The four $b$--jets were combined into two $b$--jet pairs and only events where the invariant mass of each of these pairs lied within a 45\,GeV window around $m_A$ and satisfied the asymmetry cut $\bar{\alpha} < 0.2$ were selected. This criterion reduces the background significantly. The vector $p_T$-sum of the $b$--jet pairs estimated the $p_T$ of the reconstructed $A$s, which are identified as $A_1$ and $A_2$ such that $p_T(A_1) > p_T(A_2)$ (since $A_2$ originates from the $H^\pm$ decay and is softer).

\item We calculated the invariant mass of the $2b+jj$ system, where `$2b$' is the softer pair (identified as the $A_2$), to obtain the $m_{\hpm}$. 

\item When more than one pairings of the $b$--jets satisfy the above condition, we selected the combination which maximized the separation $\Delta R = \sqrt{(\Delta\eta)^2+(\Delta\phi)^2}$ of the reconstructed $H^\pm$ and $A_1$.

\end{enumerate}


\subsection{Reconstruction of the $H$} \label{sec:h-rec}
\begin{enumerate}[leftmargin=*] 
\itemsep-0.05cm

\item The dominant signal process is $q\bar{q} \to A_1 H \to A_1 A_2 A_3 \to 4b+X$, so each event should contain at least six $b$--tagged jets. We combined these into three $b$--jet pairs and selected the combination for which the invariant mass of each pair lied within a 45\,GeV window around $m_A$, and also satisfied the $\bar{\alpha}$-cut.\footnote{We note here that the reconstruction efficiency for all the Higgs bosons can be further improved by imposing $\bar{\alpha} < 0.1$ and some other selection criteria used in \cite{CMS:2022usq}. However, due to the large cross section of the QCD background, simulating it for such a strong selection cut would require much more substantial computational resources.}

\item The $p_T$ of each $b$--jet pair was obtained by summing the 4--momenta of the two $b$--jets in it. Out of the three pairs, we identified the one with the highest $p_T$ as the prompt $A_1$. The remaining system of 4 $b$--jets then corresponded to the $A_2A_3$ pair from $H$ decay, and its invariant mass thus reconstructed the $m_H$.

\item As in the case of the $H^\pm$, if multiple pairings of the $b$--jets satisfied the above criteria, we used that $4b$--jet system for reconstructing the $H$ which maximized its separation from the third $b$--jet pair (i.e., the prompt $A_1$) in the $\{\eta,\,\phi\}$ space.

\item Since tagging 6 $b$--jets is highly challenging due to finite (mis-)tagging, events with at least 5 $b$--jets were also used for reconstructing the $H$. In this case, the light jet with the leading $p_T$ was first assumed to be the 6th $b$--jet for performing steps 1 -- 3. If this jet failed to satisfy the pairing criteria above, these steps were repeated sequentially for the jet with the next highest $p_T$, until the correct jet was found.  

\end{enumerate}


\section{Significances at the LHC}
\label{sec:signi}

\begin{figure}[t!]
  \includegraphics[width=8cm]{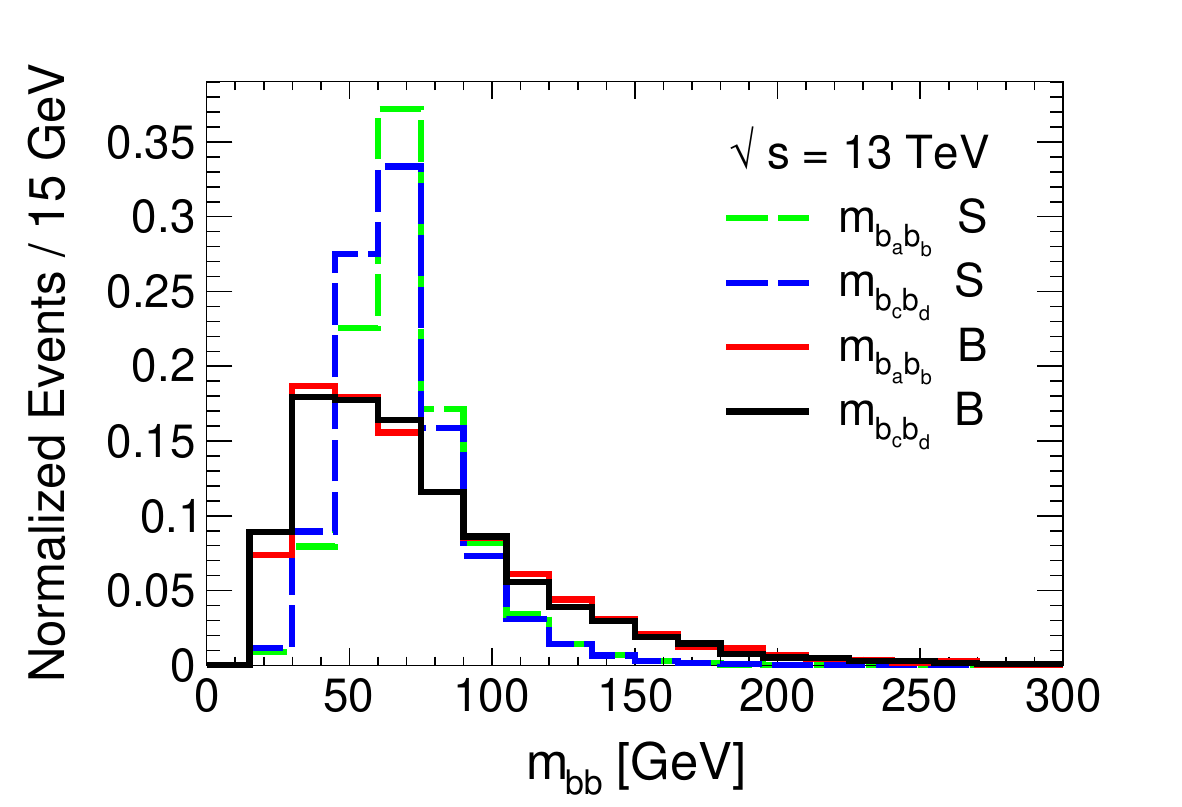}
  \caption{$m_{bb}$ distributions for the signal (green/blue - dashed) and background (red/black - solid) events for the BP.}
  \label{fig:norm-m-bb}
\end{figure}

\begin{figure}[t!]
  \includegraphics[width=8cm]{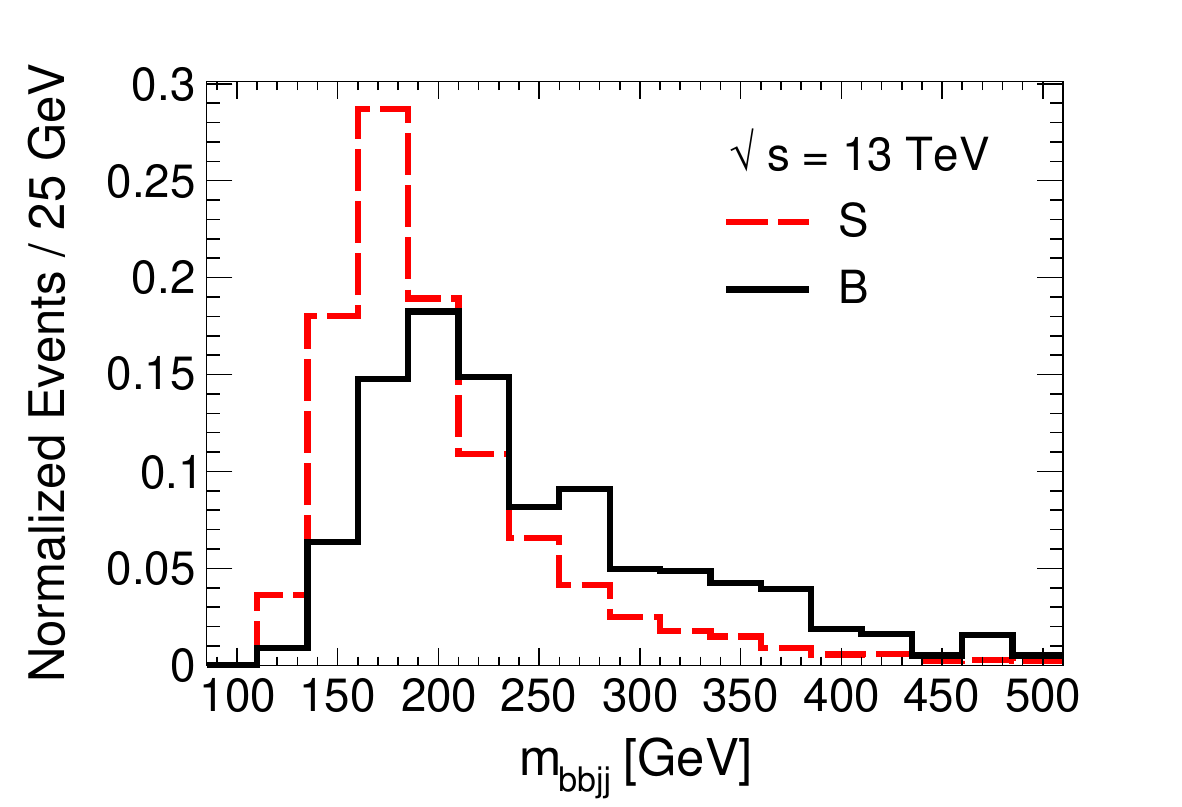}
  \caption{$m_{bbjj}$ distributions for the signal (red - dashed) and background (black - solid) events for the BP.}
  \label{fig:norm-m-bbjj}
\end{figure}

Using the steps detailed in the previous section, we calculated the signal (background) event rates, $S$ ($B$) assuming an integrated luminosity of 3000\,fb$^{-1}$ at the LHC for our BP. In Fig.~\ref{fig:norm-m-bb} we show the normalized invariant mass distributions of the $b$-jet pairs for these events. The subscript $a$ implies the distribution for the pair containing the leading $b$--jet. The signal distributions in this figure as well as the subsequent figures include all the signal modes mentioned in Sec. \ref{subsec:AA}, while the background distributions include both multi-jet and $t \bar{t} + {\rm jets}$. Clearly, the invariant masses peak at the true $m_A$. Fig.~\ref{fig:norm-m-bbjj} similarly shows the distributions of the $bbjj$ invariant mass, which peaks around the true $m_{H^\pm}=169.7$\,GeV. 

\begin{figure}[t!]
  \includegraphics[width=8cm]{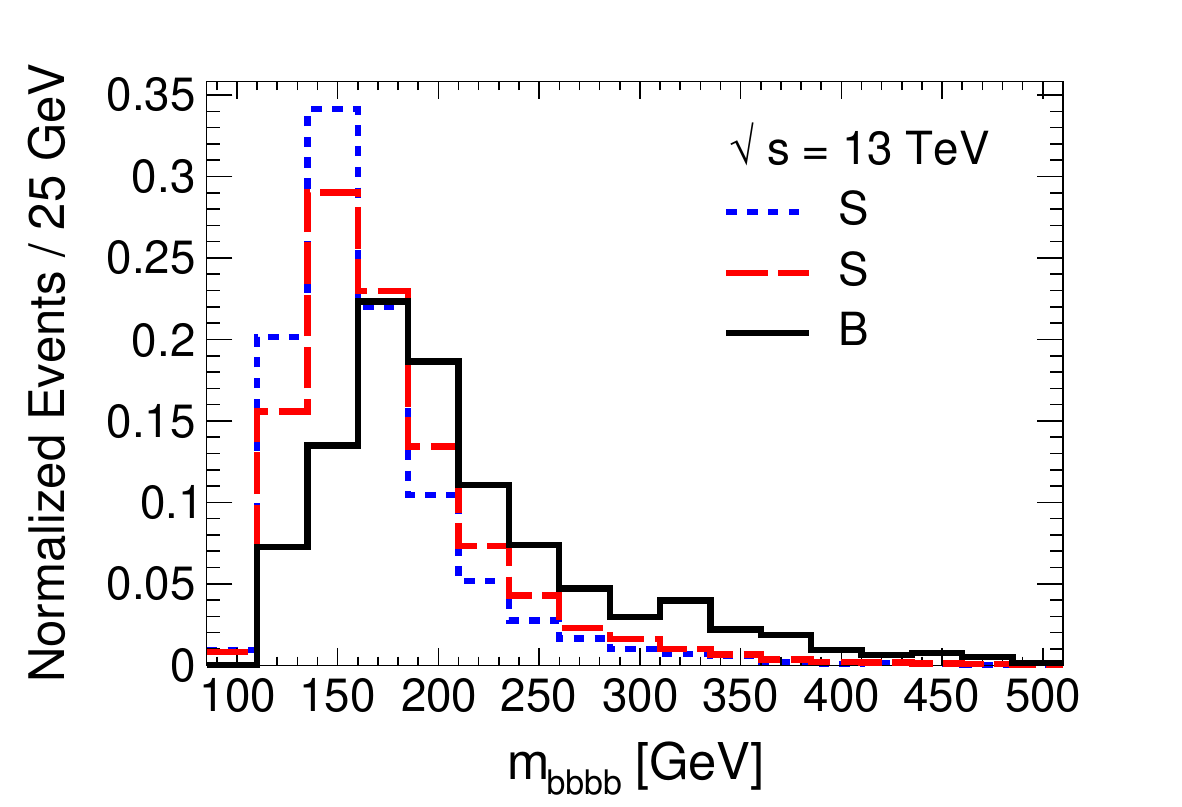}
  \caption{$m_{bbbb}$ distributions for the signal (red/blue - dashed) and background (black - solid) events for the BP.}
  \label{fig:norm-m-bbbb}
\end{figure}

Fig.~\ref{fig:norm-m-bbbb} depicts the reconstruction of the $H$, as described in Sec.~\ref{sec:h-rec}. The red-dashed signal histogram, corresponding to events with at least 5 $b$--jets, has a peak around the true $m_H=144.7$\,GeV. In this figure, the blue-dotted histogram shows the invariant mass distribution when events with 6 $b$--tagged jets are considered, which results in a better mass reconstruction compared to events with 5 $b$--jets. However, as noted earlier, estimation of the background for events with 6 $b$--jets is beyond the reach of our analysis. 

From these histograms, we chose three bins around the mass of each of the non-SM Higgs boson to estimate the statistical significance, $S/\sqrt{B}$ of its signature. For the reconstruction of the $A$, the $S$ ($B$) implies the mean of the number of events in the bins covering $m_{bb}$ from 45\,GeV to 90\,GeV for the two signal (background) distributions in Fig.~\ref{fig:norm-m-bb}. 

These significances are shown in Table~\ref{tab:params}. The highest significance was obtained for the $A$, since all the signal modes contribute to its reconstruction. In the case of the $H^\pm$ and $H$ instead, the reconstruction algorithms are based on the signal topologies $AAW$ and $AAA$, respectively, and the other contributions thus get diminished. For the $H$, the requirement of at least 5 $b$--tagged jets strongly suppresses the QCD background compared to the signal within the three relevant invariant mass bins, which leads to a considerably higher $S/\sqrt{B}$ for it compared to the $H^\pm$, as seen in the table. We point out again that, for the $H$ signal, this significance has been calculated for 5 $b$--tagged jets and one light jet (rather than for 6 $b$--jets). 

\begin{table}[tbp]
\centering
\begin{tabular}{|c|c|c|c|c|}
\hline
\multirow{2}{*}{Parameters} & Pre-selection & \multicolumn{3}{c|}{Reconstructed Higgs}\\ 
& cross section (fb) & $\sigma_S$ (fb) & $\sigma_B$ (fb) & $S/\sqrt{B}$ \\ \hline\hline
$m_{H^\pm}=169.7$ & $AAA$: 171.6 & \multicolumn{3}{c|}{$A$} \\ \hline
$m_{H}=144.7$ & $AAZ$: 0.76 & 15.4 & 8864 & 8.9$\sigma$ \\ \hline
$t_\beta = 7.47$ & $AAWW$: 25.2 & \multicolumn{3}{c|}{$H^\pm$} \\ \hline
$s_{\beta - \alpha}=0.99$ & $AAW$: 142.3 & 2.22 & 482 & 5.5$\sigma$ \\ \hline
$m_{12}^2=2355$ & $AAAW$: 79.7 & \multicolumn{3}{c|}{$H$} \\ \hline
BR$(AA)=0.99$ & $AAZW$: 0.35 & 2.55 & 309 & 7.9$\sigma$ \\
\hline
\end{tabular}
\caption{Column 1: Input parameter values and BR($H\to AA$) for the BP (all masses are in GeV). Column 2: Cross section for each of the signal channels, assuming a next-to-next-to-leading order $k$-factor of 1.35 \cite{Bahl:2021str}. Columns 3--5: Total signal and background cross sections after applying all the selection cuts, and the discovery significance for the three non-SM Higgs bosons.}
\label{tab:params}
\end{table}

\section{Conclusions}
\label{sec:summa}

In a new physics framework containing multiple Higgs fields, such as the 2HDM studied here, a full reconstruction of the Higgs potential would entail observing all the additional physical Higgs states and measuring their masses and couplings. 
Numerous attempts, both theoretical and experimental ones, have been made to extract signatures of the two neutral Higgs bosons, besides the SM-like one, as well as the charged scalar in various Types of the 2HDM. These studies, however, generally focus on a QCD-induced single- or multiple-production, followed by a specific decay channel, of any one of these additional states, for investigating its discovery prospects at the LHC. 

In this study we have shown, for the very first time, that all the three non-SM Higgs bosons in this model might be detectable in the unique final state with 4 (or more) $b$--jets at the HL-LHC. This is possible for specific (and rather narrow) parameter space configurations, wherein intermediate pairs of relatively light Higgs bosons, produced on-shell, lead to multi-$A$ states, which subsequently decay in the $b\bar{b}$ channel. Our sophisticated MC analysis yielded a $S/\sqrt{B}>5\sigma$ for the signals of all the three non-SM Higgs states.

We therefore strongly advocate systematic investigations of the EW-induced processes alongside the time-honoured QCD-initiated ones, as they may prove crucial for nailing down the Type-I 2HDM as (the low-energy limit of) the new physics framework prevalent in nature.  

\begin{acknowledgments} 

SMo is supported in part through the NExT Institute and the STFC Consolidated Grant ST/L000296/1. PS is supported by the National Research Foundation of Korea, Grant No. NRF-2022R1A2C1007583. TM's work is supported by JSPS KAKENHI Grant Number 22F21324. The authors would also like to thank the KIAS Center for Advanced Computation for providing computing resources.
\end{acknowledgments} 

%

\section*{Appendix: A lighter $A$}

\begin{figure}
\begin{tabular}{cc}
\includegraphics[width=9cm]{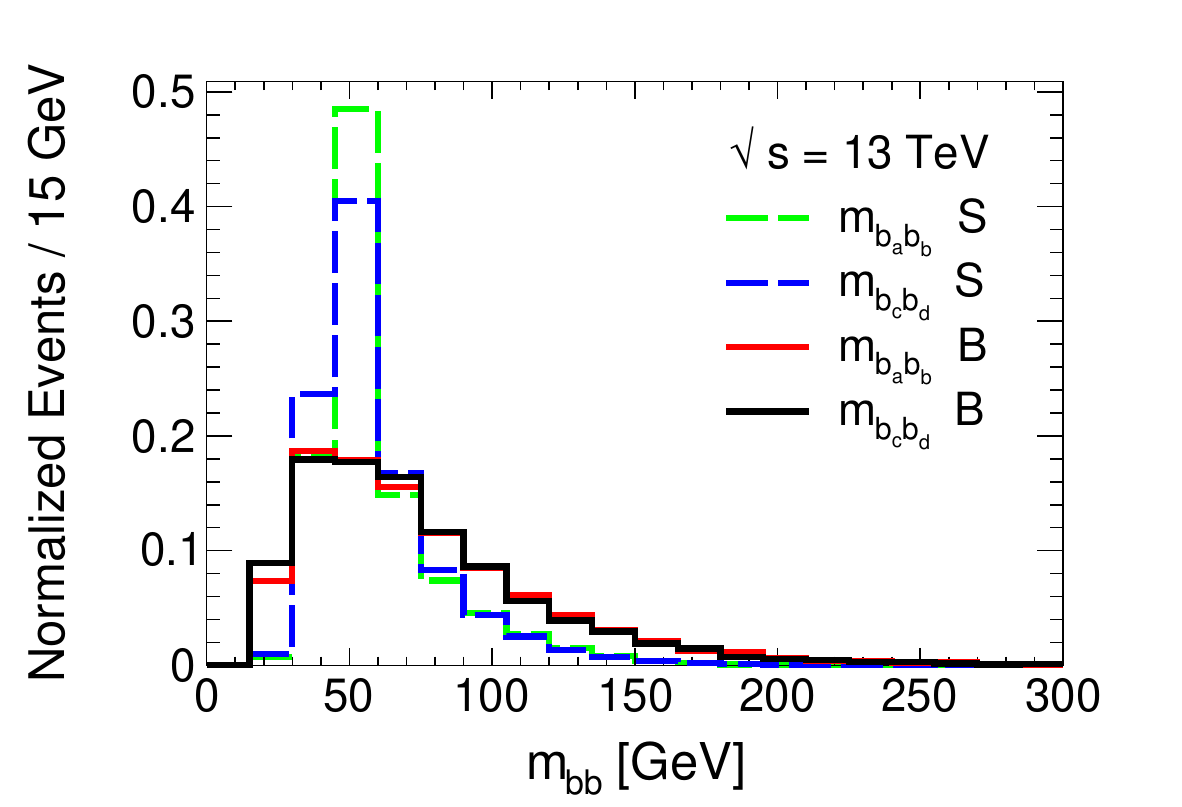}
&
\includegraphics[width=9cm]{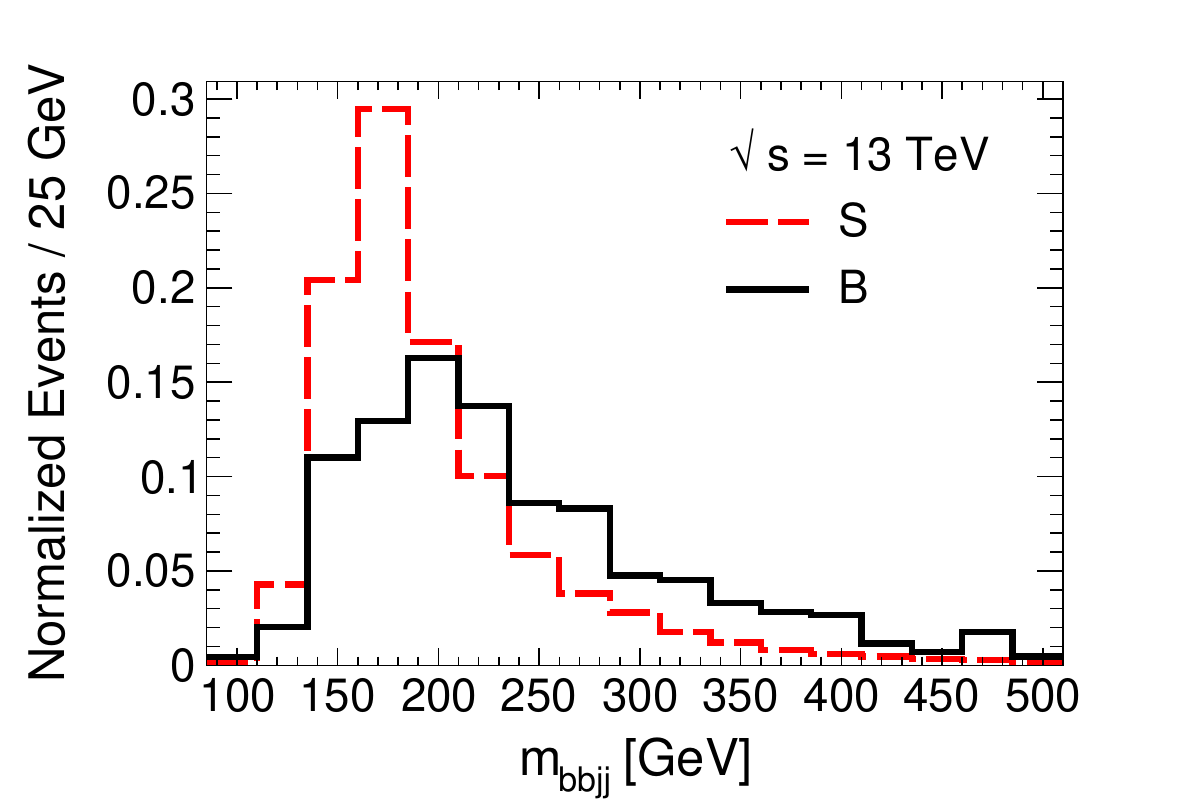}
\\
\multicolumn{2}{c}{\includegraphics[width=9cm]
{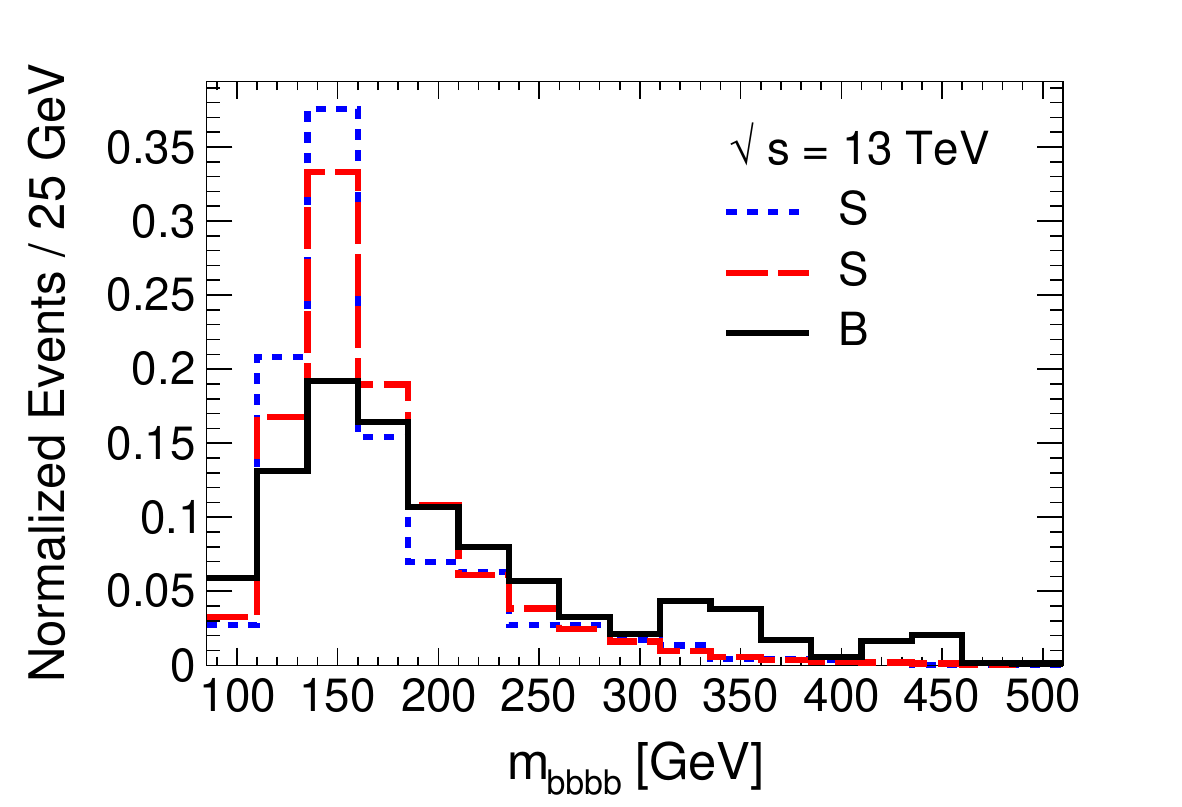}}
\end{tabular}
\caption{\label{fig:BPPdist} Normalized invariant mass  distributions for the BP$^\prime$. Top-left: $m_{bb}$ for the signal (green/blue - dashed) and background (red/black - solid) events. Top-right: $m_{bbjj}$ for the signal (red - dashed) and background (black - solid) events. Bottom: $m_{bbbb}$ for the signal (red/blue - dashed) and background (black - solid).}
\end{figure}

In order to test the efficiency of our reconstruction method for the scenario with substantial partial width of the decay of the $H$ into $AZ$, we picked a BP$^\prime$, for which BR$(H\to AZ)=0.5$ and $m_A=50$\,GeV. $m_H$ and $m_{H^\pm}$ for this BP$^\prime$ are almost identical to the respective ones for our main BP, and it therefore allows us to assess the impact of also a smaller $m_A$, besides a significantly reduced BR$(H\to AA)$, on our analysis. The invariant mass distributions for the BP$^\prime$ are shown in Fig. \ref{fig:BPPdist} for the $A$ (top-left), $H^\pm$ (top-right), and $H$ (bottom), and closely resemble the respective ones for the $m_A=70$\,GeV BP.

For the BP$^\prime$, we see relatively low significances for all the Higgs bosons in Table \ref{tab:paramsBPP}. Our reconstruction algorithm for the $H$ is thus much more efficient when its decay to $AA$ is highly dominant. Furthermore, a lighter $A$ results in much softer $b$--jets, which lowers the selection efficiency overall. Despite all these deficiencies, the signal significances are still a formidable $>3\sigma$ for all the Higgs bosons for this BP$^\prime$, thus demonstrating the strength of our proposed reconstruction method.

\begin{table}[h]
\centering
\begin{tabular}{|c|c|c|c|c|}
\hline
\multirow{2}{*}{Parameters} & Pre-selection & \multicolumn{3}{c|}{Reconstructed Higgs}\\ 
& cross section (fb) & $\sigma_S$ (fb) & $\sigma_B$ (fb) & $S/\sqrt{B}$ \\ \hline\hline
$m_{H^\pm}=169.8$ & $AAA$: 101.3 & \multicolumn{3}{c|}{$A$} \\ \hline
$m_{H}=150.0$ & $AAZ$: 79.3 & 10.4 & 10175 & 5.7$\sigma$ \\ \hline
$t_\beta = 17.1$ & $AAWW$: 27.7 & \multicolumn{3}{c|}{$H^\pm$} \\ \hline
$s_{\beta - \alpha}=0.98$ & $AAW$: 198.0 & 1.33 & 491 & 3.3$\sigma$ \\ \hline
$m_{12}^2=1275$ & $AAAW$: 37.1 & \multicolumn{3}{c|}{$H$} \\ \hline
BR$(AA)=0.48$ & $AAZW$: 29.0 & 1.06 & 256 & 3.6$\sigma$ \\
\hline
\end{tabular}
\caption{Input parameters, signal and background cross sections, and discovery significances for the three non-SM Higgs bosons of the BP$^\prime$.}
\label{tab:paramsBPP}
\end{table}

\onecolumngrid

\twocolumngrid

\bibliography{2HDM1_v02}

\end{document}